\documentclass[aps,prl,showpacs,showkeys,twocolumn]{revtex4-1}
\usepackage{graphicx,hyperref,amsmath,amsfonts}
\usepackage{epstopdf,color,bm,multirow,rotating,soul}

\begin{document}
\title{Broadband Multiresonator Quantum Memory}
\author{S.A. Moiseev$^{1,2 *}$, F.F. Gubaidullin$^{2}$, R.S. Kirillov$^{3}$, R.R. Latypov$^{3}$, N.S. Perminov$^{1,2}$, K.V. Petrovnin$^{3}$, O.N. Sherstyukov$^{3}$}
\affiliation{$^{1}$Kazan Quantum Center, Kazan National Research Technical University n.a. A.N.Tupolev-KAI, 10 K. Marx, Kazan 420111, Russia}
\affiliation{$^{2}$Zavoisky Physical-Technical Institute of the Russian Academy of Sciences, 10/7 Sibirsky Tract,
Kazan 420029, Russia}
\affiliation{$^{3}$Kazan Federal University, 18 Kremlyovskaya Str., Kazan 420008, Russia}
\email{samoi@yandex.ru}

\begin{abstract}
In this paper we present universal broadband multi-resonator quantum memory (MR-QM) based on the spatial-frequency combs of the microresonators coupled with a common waveguide.
Here we found new Bragg-type impedance matching condition for the coupling of the microresonators with a waveguide field which provides an  efficient broadband quantum storage. 
The obtained analytical solution for the  microresonator fields enables sustainable parametric control of all the memory characteristics.
\end{abstract}

\maketitle

\section{I. Introduction}
Control and manipulation of  electromagnetic fields play a basic role in quantum information technologies \cite{Kurizki2015}.
In the last decade these matters attracted a new wave of interest among those working on quantum storage of single photon fields used as quantum information carriers \cite{Lvovsky2009,Hammerer2010, Simon2010}.
Herein, quantum memory (QM) devices being experimentally elaborated seem to be the key elements of the quantum computer \cite{Grezes2015,Mariantoni2011,Hill2015} and repeater \cite{Sangouard2011} of the long-distance quantum communication.
The main stream in current  theoretical and experimental works on the QM creation is based on the usage of the resonant atomic ensembles as the robust keepers of large amount of quantum data. 
The use of solid-state media, especially ensembles of the electron-nuclei spin systems, will be very promising for the long-lived quantum storage creation. 
The rare-earth ions doped in inorganic crystals \cite{Tittel2009,Zhong2015}, NV-centers in diamonds \cite{Fuchs2011,Lukin2012}, $P$ donors in an isotopically pure $Si$ crystal \cite{Morton2008} can serve as such media, particularly for the use in microwave quantum storage \cite{Kubo2011,Probst2013,Julsgaard2013,Gerasimov2014,Grezes2014,Grezes2015,Krimer2016}.
 
The use of the photon echo approach \cite{Moiseev2001,Tittel2009} with the atomic system being in the high-Q single mode resonator is considered especially promising for the practical implementation of the multi-mode quantum storage. 
This happens due to the impedance matching enhancement of the resonant interaction of a weak signal light field with a small number of atoms \cite{Afzelius2010,Moiseev2010,Sabooni2013_1,Sabooni2013_2,Jobez2014,Thierry2014}. 
However, this impedance matching scheme has a limited spectral range,  since its range is determined by the quality factor of the resonator \cite{Moiseev2010,Moiseev2013_1,EMoiseev2016}. 

\begin{figure}[h]
\includegraphics[width=0.40\textwidth]{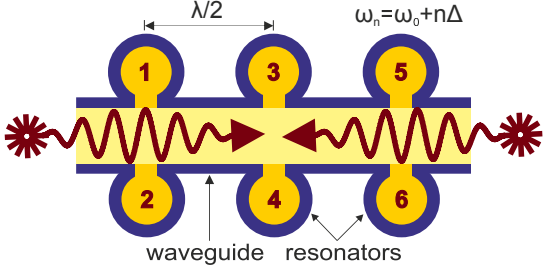}
\caption{(Color online) Spatial scheme of MR-QM; $(1)-(6)$ are single mode microwave microresonators  with distance $\lambda/2$ between the nearest ones; the microresonators contain dielectric insertions which ensure the smallness of their spatial sizes in comparison with $\lambda$; wavy lines  represent the propagating  electromagnetic fields interacting with the microresonators.}
\label{general_scheme}
\end{figure}

In this paper, we propose a new approach providing broadband impedance matching QM.
Instead of atomic ensembles (electron spins etc) in a single mode resonator, in this approach (we call it Multi-Resonator QM or MR-QM approach), we use an array of high-Q single mode microresonators located at the distance of a whole number of half wavelengths from each other and coupled to a broadband waveguide.
The microresonator frequencies form a periodic structure which covers the broadband frequency range. 
The principle spatial scheme of the MR-QM is depicted in Fig. \ref{general_scheme}.     

It is worth noting that waveguide structures in which many microresonators are connected to the common waveguide \cite{Foresi1997} have been intensively studied in optics  \cite{Boriskina2010,Morichetti2012,Bogaerts2012}.
Great attention was paid to the structures consisting of the microresonators with equal frequencies. 
In this case the realized nanooptical structures have demonstrated promising properties for the on-chip optical time delay lines \cite{Xia2007}.
We would also like to highlight various schemes in which the optical waveguides are coupled with microresonators characterized by the frequencies that are tunable within a wide spectral range \cite{Saperstein2007} by using the on-chip comb filters based on Bragg gratings (LCFBG)\cite{Giuntoni2011,Pruessner2007_1,Pruessner2007}, microring resonators \cite{Dong2007,Lee2008}, and cascaded Sagnac loop mirrors (SLMs) \cite{Sun2013,Jiang2016}(see also recent reviews \cite{Heck2013,Du2016}).
Here, silica \cite{Wang2016} and crystalline microresonators \cite{Ilchenko2004,Grudinin2015} are the most encouraging for the quantum processing purposes due to their higher quality factors. 
These optical integrated waveguide structures may also be useful for the implementation of the proposed MR-QM. 

We show that the MR-scheme can be interesting for the universal quantum storage of  light field in a broadband frequency range.
Here, we have found that the perfect storage is possible at the optimal parameters of the microresonators and their coupling with the waveguide. 
The optimal parameters are determined by a single new impedance matching condition for the interaction of free propagating light field with the microresonators due to the fact that they form a spatial Bragg-grating. 
We decided to call it Bragg-type impedance matching condition.

\section{II. Physical model}

In this part we give a theoretical study of the proposed QM scheme and analyze its basic properties.

\subsection{Theoretical analysis}
To simulate the interaction of light field with array of nearby microresonators in the broadband waveguide, we use the following multi-particle Tavis-Jaynes-Cummings type model \cite{Jaynes1963,Tavis1968} with the Hamiltonian:

\begin{align}{\label{hamilt}}
& \nonumber \hat{H}=\int dkc|k|\hat{b}_k^{\dagger}\hat{b}_k+\sum_n[ck_0+n\Delta ]\hat{a}_n^{\dagger}\hat{a}_n\\
& +g\int dk\sum_n\left[e^{-ikzn}\hat{b}_k^{\dagger}\hat{a}_n+e^{ikzn}\hat{a}_n^{\dagger}\hat{b}_k\right],\end{align}

\noindent
where $c$ is a speed of light, integer index $n\in\{1,...,N\}$ corresponds to $n$-th microresonator, continuous index $k\in[-k_0-\delta_0;-k_0+\delta_0]\cup[k_0-\delta_0;k_0+\delta_0]$ corresponds to the forward and backward waveguide field modes located near the resonant frequency $\omega_0=ck_0$; $\hat{a}_n,\hat{a}_n^{\dagger}$ and $\hat{b}_k,\hat{b}_k^{\dagger}$ are the annihilation and creation Bose-operators of microresonator modes and of waveguide field modes, $z_n=zn$ is a spatial coordinate of $n$-th microresonator, where $n$-th microresonator has the frequency of $\omega_n=\omega_0+\Delta n$ and $\Delta$ is a spectral distance between the nearest microresonator frequencies. 

The first term in (\ref{hamilt}) corresponds to the energy (in the units $\hbar$) of the waveguide modes and the second term describes the energy of $N$ microresonators. 
The third term in (\ref{hamilt}) corresponds to the interaction of the waveguide field with the microresonator modes which is characterized by  the coupling constant $g$. 
Point location of microresonators in (\ref{hamilt}) is provided by their contact localized coupling with a waveguide field as it is shown in Fig. \ref{general_scheme}. 
We study the interaction of a single photon field with the analyzed compound MR-system in the broadband waveguide.  
By taking into account the Hamiltonian (\ref{hamilt}), the wave function of this compound system  has the following form:
\begin{align}\label{wave_func}
|\psi\rangle=\left[\int dkf_k(t)\hat{b}_k^{\dagger}+\sum_n\alpha_n(t)\hat{a}_n^{\dagger}\right ]|0\rangle,
\end{align}
where $f_k(t)$ and $\alpha_n(t)$ are the amplitudes of waveguide and  microresonator field modes which satisfy the normalization condition $\int dk~f_k^*(t)f_k(t)+\sum_n\alpha_n^*(t)\alpha_n(t)=1$.

By using the Schr\"{o}dinger equation $(\partial_t+i\hat{H})|\psi\rangle=0$, we get the following equations for the field amplitudes $f_k(t)$ and $\alpha_n(t)$:

\begin{align}\label{dyn_eq_k_repr} 
& \nonumber {\big[} \partial_{t}+ic|k|{\big]}f_k(t)+ig\sum_ne^{-ikzn}\alpha_n(t)=0,\\
& {\big[}\partial_{t}+i\omega_0+i\Delta n{\big]}\alpha_n(t)+ig\int~dke^{ikzn}f_k(t)=0.
\end{align}
Basic properties of MR-QM can be easily studied for a retrieval stage.
In this case, we assume that the waveguide mode stays in the vacuum state at time moment $t=0$ when the MR-system have absorbed an input single photon field. 
This moment of time corresponds to the middle point between the absorption and retrieval stages.
By substituting the formal solution \\ $f_k(t)=-ig\int\limits_{0}^{t}d\tau~\sum_m e^{-i|k|[c(t-\tau)+\operatorname{sign}(k)zm]}\alpha_m(\tau)$ in the second Eq. (\ref{dyn_eq_k_repr}) and by replacing variables $\alpha_n(t)=e^{-i\omega_0t}\beta_n(t)$, we get:

\begin{align}\label{dyn_eq_beta_retr}
& \nonumber [\partial_t+i\Delta n]\beta_n(t)\\
& +\frac{\pi g^2}{c}\sum_{m}e^{ik_0z|n-m|}\beta_m(t-|n-m|z/c)=0,
\end{align}
where we used the following initial condition: $f_{k}(0)=0,~\beta_n(0)=c_n,~t\in[0;2\pi/\Delta]$. 
In case of a single microresonator ($n=0$), we find the solution $\beta_0(t)=\operatorname{exp}\{-\Gamma t/2\}$ where  $\Gamma=2\pi g^2/c$ determines the mode linewidth.
For a sufficiently narrow spectral width of an irradiated light pulse $|(n-m)z/c|\ll\delta t_{s}$ (where $t_{s}$ is a typical temporal duration of the pulse) and periodic spatial positions of microresonators $\left\{z=l\pi/k_0,~l\in\mathbb{Z}\right\}$, we obtain the following solution:
\begin{align}\label{beta_retr_large_N}
& \nonumber \beta_n(t)=e^{-i\Delta nt}\left[c_n-\frac{\Gamma}{2\Delta}\left[1+\frac{\pi\Gamma}{2\Delta}\right]^{-1}\right.\\
& \left. \times\sum_m\frac{\sin(\Delta (n-m)t/2)}{(n-m)/2}e^{i\Delta (n-m)t/2}c_m\right],
\end{align} 
where we introduced new variables: $(-1)^{ln}\beta_n(t)\rightarrow \beta_n(t),~(-1)^{ln}c_n\rightarrow c_n$. 
  
Solution (\ref{beta_retr_large_N}) demonstrates quite  complex dynamics of the microresonator fields for arbitrary parameters of the interaction.  
Further, from Eq. (\ref{beta_retr_large_N}) we find the quantum efficiency $\eta(t)=1-\sum_n|\beta_n(t)|^2$ of the field irradiation as the following product of two factors: 

\begin{align}\label{ef_cont_retr}
& \eta(t)=\eta_0(g)\eta_1(t),\nonumber \\
& \eta_0(g)=\frac{2\pi\Gamma}{\Delta}\left[1+\frac{\pi\Gamma}{2\Delta}\right]^{-2},\nonumber \\
& \eta_1(t)=\int\limits_{0}^{\Delta t/(2\pi)}dv\left|\sum_ne^{-i2\pi nv}c_n\right|^2.
\end{align}

\noindent

Firstly, we note that the solution (\ref{ef_cont_retr}) is a time reversible and quantum efficiency of the entire process is given by $\eta_{total}=\eta^2(T)$ (where $T=2\pi/\Delta$). 
In this case, the temporal evolution of the system under consideration describes a storage of the input light pulse at the absorption stage (i.e for $t<0$). 
Secondly, as it is seen in Eq. (\ref{beta_retr_large_N}),   the input and output light pulses will be identical to each other for the symmetric initial conditions $c^*_n=c_{-n}$ which means a perfect fidelity for the retrieval of the input pulse. 
These two findings complete the analytic description of the studied MR-QM scheme. 
The solutions  (\ref{beta_retr_large_N}) and  (\ref{ef_cont_retr}) are the main analytic results of our work. 

The studied quantum storage is valid  for the light pulses with spectral width up to $\sim N\Delta$ which can be much  larger in comparison with the linewidth $\Gamma$ of a single microresonator mode.
Below we analyze a condition for realization of highly efficient quantum storage.

\subsection{Optimum condition and universality}

The general properties and role of the impedance matching condition  are well-known in laser physics \cite{Haus1984,Siegman1986}.
Lately  such condition became important for the QM realization on the atomic ensembles in a single mode resonator  \cite{Afzelius2010,Moiseev2010,Moiseev2013_1,EMoiseev2016} and for controlling the perfect light-atoms quantum dynamics  \cite{Kalachev2013,Gerasimov2014,Sokolov2015} in similar systems.
The impedance condition physically adjusts the optimal interference effects of the light media interaction in the resonant cavities that significantly facilitates the implementation of the main requirements to the quantum dynamics of light.
First experiments of the impedance matched photon echo QM schemes have been successfully demonstrated  in the optical cavities   \cite{Sabooni2013_1,Sabooni2013_2,Jobez2014}.

By using Eq.(\ref{ef_cont_retr}) we can analyze a condition of the efficient light field irradiation. 
The temporal evolution of the irradiation is governed by the factor $\eta_1(t)$, which can be close to unity at time moment $t =2\pi/\Delta $  as it is discussed in the next section.
From the factor $\eta_0(g)$, we find that the maximum quantum efficiency (i.e. $\eta(T)=1$) is reached under the following condition for the coupling constant $g$ (or linewidth $\Gamma$ of the microresonator mode) and spectral distance $\Delta$ between the microresonator frequencies:
\begin{align}{\label{extr_cond_cont_retr}}
& \Delta=\frac{\pi}{2}\Gamma,
\end{align}
we call it Bragg-type impedance matching condition providing the perfect MR-QM operation.
\begin{figure}[h]
\includegraphics[width=0.4\textwidth]{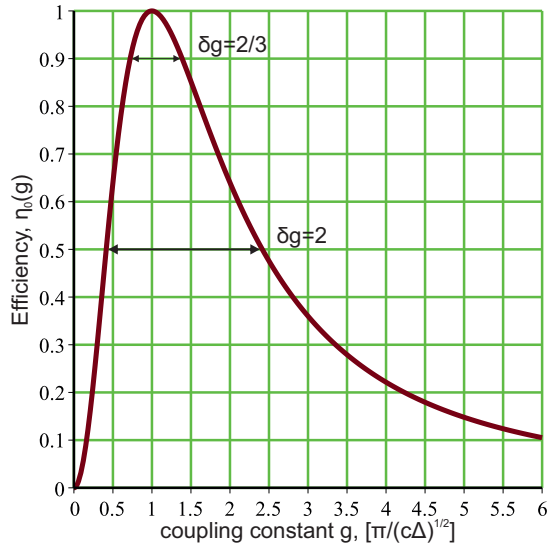}
\caption{(Color online) Quantum efficiency $\eta_0=4\tilde{g}^2/(1+\tilde{g}^2)^2$ vs the dimensionless coupling constant $\tilde{g}=[\pi\Gamma/(2\Delta)]^{1/2}$; the maximum efficiency  $\tilde{g}=1$ corresponds to the Bragg-impedance matching condition.}
\label{eta_0_g}
\end{figure}
As it is seen in Fig. \ref{eta_0_g}, the quantum efficiency clearly depends on the dimensionless coupling constant $\tilde{g}=[\pi\Gamma/(2\Delta)]^{1/2}=g\pi/(c\Delta)^{1/2}$.
We can control the values of $g$ and $\Delta$ which facilitates the implementation of the condition (\ref{extr_cond_cont_retr}). 
We also see in Fig. \ref{eta_0_g} that the efficiency $\eta_0(g)$ is higher than $90$\% in the interval $\tilde{g}\in[(\sqrt{10}-1)/3,(1+\sqrt{10})/3]$ in which it is possible to change the coupling constant in the range $\delta \tilde{g}\leq 2/3$.
The variations of $g$ and $\Delta$ show a stability of high efficiency. 
In turn the fulfillment of the requirement $\eta_1(2\pi/\Delta)=1$ for all the initial conditions of the microresonator states means that the MR-QM is universal within some spectral range.

For comparison, we also note that the usual impedance matching scheme based on using the atomic ensemble in a single mode high-Q cavity  \cite{Afzelius2010,Moiseev2010,Sabooni2013_1,Sabooni2013_2,Jobez2014,Thierry2014} requires at least two impedance conditions (one of which is a spectral condition) for the quantum storage in a spectral width comparable with the linewidth of an empty cavity, as it was shown in a series of works\cite{Moiseev2010, Moiseev2013_1,EMoiseev2016}.
However, it is sufficient to satisfy the only one Bragg-type impedance condition (\ref{extr_cond_cont_retr}) for implementation of perfect QM within even broader spectral range in the considered scheme.

It is worth noting that the first scheme with periodic spatial-frequency structure of resonant atomic lines was considered for the optical QM in \cite{Tian2013} before. 
However, we have found that the point-like spatial grating scheme can demonstrate the Bragg-type impedance matching condition for the quantum storage in the broadband waveguide.
That is, the  impedance matching condition  becomes possible  due to the point-like coupling of the light field with the microresonators forming a spatial grating  (see Fig. \ref{general_scheme}).
Herein, the spatial grating of the microresonators provides a role of a broadband distributed resonator for the light fields. 
Thus, any  use of additional (spectral) impedance matching condition is not necessary.

\subsection{MR-QM dynamics}
The MR-scheme has a unique quality: simple analytical solution (\ref{beta_retr_large_N}) of the microresonator fields enables a sustainable parametric control of all the memory characteristics. 
However, the detailed analysis of the dynamic structure of the QM-schemes is a laborious fundamental problem for all multi-particle quantum systems, and a general approach to this problem for the studied MR-scheme will be presented in future investigations. 

\begin{figure}[h]
\includegraphics[width=0.4\textwidth]{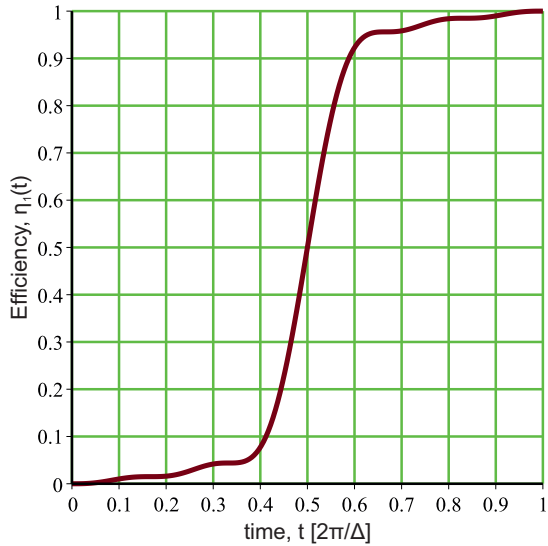}
\caption{(Color online) The current efficiency $\eta_{1}(t)$ on time t  in units $2\pi/\Delta$  for $N=6$. Time interval $t\in[0,\pi/\Delta]$ corresponds to half the storage phase.}
\label{efficiency_1}
\end{figure}

Primarily, we focus on the important case of a short rectangular signal pulse when the initial data satisfy the condition $c_n=(-1)^nN^{-1/2}$ and the following relation 

\begin{align}{\label{spin_corr_case_retr}}
& \nonumber \eta_{1}(2\pi u/\Delta)=\\
& \frac{1}{N}\int\limits_{0}^{u}dv\left[\frac{\sin[N\pi(v-1/2)]}{\sin[\pi(v-1/2)]}\right]^2\xrightarrow{N\rightarrow\infty}\theta(u-1/2),
\end{align}

\noindent
is received for the factor $\eta_{1}(2\pi u/\Delta)$ of quantum efficiency (\ref{ef_cont_retr}), where $\theta(u)$ is a Heaviside step function ($\theta(0)=1/2$).

\begin{figure}[h]
\includegraphics[width=0.40\textwidth]{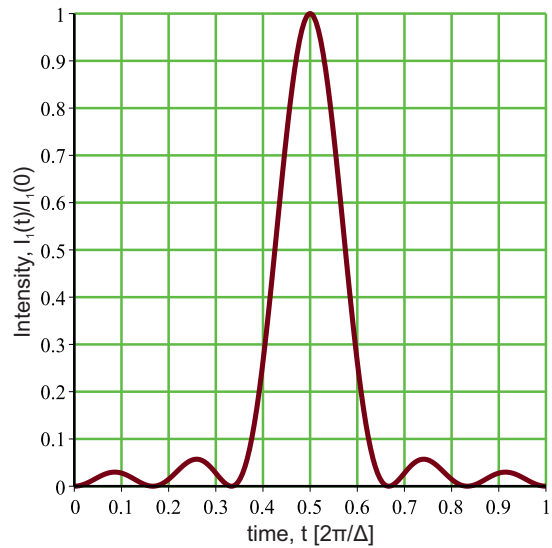}
\caption{(Color online) Normalized intensity of echo pulse emission $I_{1}(t)=\partial_t\eta_{1}(t)$ as a function of time t in units $2\pi/\Delta$  for $N=6$; the pulse duration is about $\delta t_{s}\sim 2\pi/(N\Delta)$ that allows the use of short pulses.}
\label{intensity_1}
\end{figure}

Dynamic structure of QM behavior for $N=6$ is determined by the curves in Figs. (\ref{efficiency_1}),  (\ref{intensity_1}). 
Time interval $t\in[0,\pi/\Delta]$ corresponds to half the storage phase. 
We already see a good enough light storage  for $N=6$. 
The envelope curve in the Eq. (\ref{spin_corr_case_retr}) has a typical decay of the form $e^{-N u}$ which provides  a quite perfect storage phase for any $N>6$. 
The maximum intensity $I_{1}(t)$ of the irradiated field is observed at $t=\pi/\Delta$. 
In this case, the current quantum efficiency of the field irradiation is given as $\eta_1(t)=\int_0^t dt I_1(t)/\int_0^{\infty} dt I_1(t)$ and, respectively, $I_{1}(t)=\partial_t\eta_{1}(t)$ (where we have used a normalization $\int_0^{\infty} dt I_1(t)=1$ and Bragg-type impedance matching condition (7)). 
Thus, all the energy stored initially in the microresonators will be irradiated  eventually  into the common waveguide. 
Herein, the minimum pulse duration of the signal pulses is limited by the value $\delta t_{s}\sim 2\pi/(N\Delta)$ that is determined by the total spectral width $N\Delta$ of the frequency comb structure, while shorter signal pulses will be out of resonant interaction with the MR-structure.   

\begin{figure}[h]
\includegraphics[width=0.4\textwidth]{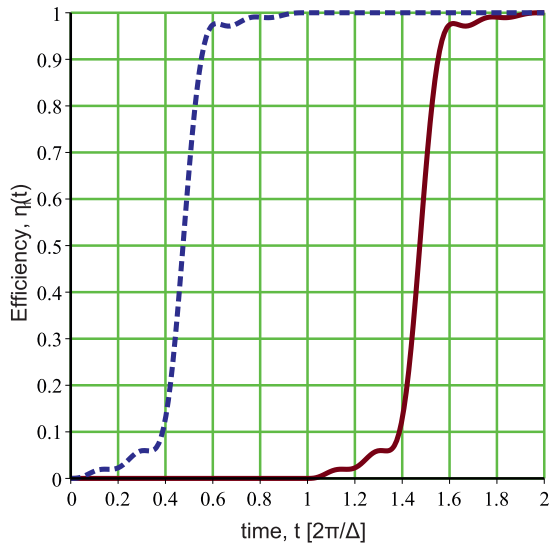}
\caption{(Color online) The current efficiencies of the echo emission $\eta_{1}(t)$ on time t in units $2\pi/\Delta$ for $N=6$ at the matching condition (\ref{extr_cond_cont_retr}): blue dotted line - free evolution, red line - evolution with increased lifetime of the stored pulse due to the alignment of the microresonator frequencies ($\Delta=0$)  during the time interval  $T=2\pi/\Delta$.}
\label{storage_phase}
\end{figure}

In turn, we can significantly increase the storage time by dynamic control of the spectral detunings between the microresonator modes after complete absorption of the input signal pulse.
Namely, in accordance with Eq. (\ref{beta_retr_large_N}), the evolution of each microresonator mode will be frozen for a given time interval $T$ if all the microresonator frequencies is adiabatically quickly aligned with each other (i.e. $\Delta=0$). 
Herein, the microresonator field modes will stay in the so called nonradiative dark state during this time interval.   
However, after restoring the original frequencies, the behavior of the MR-system is recovered and the emission of the recorded signal will be moved on the time interval $T$ as it is numerically demonstrated in Fig. \ref{storage_phase}  for sufficiently high intrinsic quality factor of the microresonators $Q=\omega_0/(2\gamma)$ where $\gamma T\ll 1$.

\section{IV. Conclusion}
We have studied the system of high-Q microresonators which have a point-like coupling with a broadband waveguide (MR-scheme).
The aim of these studies is to identify and demonstrate the optical conditions under which the MR-scheme can ensure high quantum efficiency during the storage and read-out of the broadband microwave fields. 
Herein, the microresonators form a spatial Bragg-grating system and a periodic structure of narrow resonance lines.
We have obtained the analytical solution (\ref{beta_retr_large_N}) describing the quantum dynamics of the microresonator fields.
Based on this solution, we have found a single new type of the impedance matching condition  (\ref{extr_cond_cont_retr}) (we called it Bragg-type impedance matching condition) providing a perfect storage and retrieval of the microwave light fields in the MR-system. 

The proposed MR-QM scheme is interesting because it allows to work with the broadband microwave fields using a system of high-Q microresonators. 
Due to the high Q-factor, it is possible to significantly enhance the interaction of the microwave fields with the individual microresonators as well as with the systems of electronic spins that could be in these microresonators. 
Such MR-scheme will greatly facilitate the practical implementation of the broadband QM-devices.
MR-scheme can be implemented employing different physical platforms: for example, using planar microwave waveguides characterized by highly reduced spatial sizes, various nanooptical waveguide structures, photonic crystals and other systems which allow to have localized interactions of many quantum objects with a broadband carrier of signals. 

\section{Acknowledgments}
The research has been supported by the Russian Government Program of Competitive Growth of Kazan Federal University and by the Russian Foundation for Basic Research through the Grant No. 15-42-02462. 

\bibliographystyle{apsrev4-1}
\bibliography{MR_QM}

\end{document}